\documentclass{aastex631}
\usepackage{amsmath}
\usepackage{amssymb}

\newcommand{\be}{\begin{equation}}
\newcommand{\ee}{\end{equation}}

\shorttitle{X-ray emission from TYC 2597-735-1}
\shortauthors{G\"unther et al.}

\graphicspath{{./}{figures/}}

\begin{document}

\title{X-ray emission from candidate stellar merger remnant TYC 2597-735-1 and its Blue Ring Nebula}

\author[0000-0003-4243-2840]{Hans Moritz G{\"u}nther}
\affiliation{MIT Kavli Institute for Astrophysics and Space Research, 77 Massachusetts Avenue, Cambridge, MA 02139, USA}
\author[0000-0002-8636-3309]{Keri Hoadley}
\affiliation{The University of Iowa, Dept. of Physics \& Astronomy, Van Allen Hall, Iowa City, IA, 52242, USA}
\affiliation{California Institute of Technology, Dept. of Physics, Mathematics, and Astronomy, Cahill Center for Astronomy \& Astrophysics, Pasadena, CA 91125, USA}
\author[0000-0002-3164-9086]{Maximilian N. G{\"u}nther}
\thanks{Juan Carlos Torres Fellow, ESA Research Fellow}
\affiliation{Department of Physics, and Kavli Institute for Astrophysics and Space Research, Massachusetts Institute of Technology, 77 Massachusetts Avenue, Cambridge, MA 02139, US}
\affiliation{European Space Agency (ESA), European Space Research and Technology Centre (ESTEC), Keplerlaan 1, 2201 AZ Noordwijk, The Netherlands}
\author[0000-0002-4670-7509]{Brian D. Metzger}
\affiliation{Department of Physics and Columbia Astrophysics Laboratory, Columbia University, Pupin Hall, New York, NY 10027, USA}
\affiliation{Center for Computational Astrophysics, Flatiron Institute, 162 5th Ave, New York, NY 10010, USA}
\author[0000-0002-5094-2245]{P. C. Schneider}
\affiliation{Hamburger Sternwarte, Gojenbergsweg 112, 21029 Hamburg, Germany}
\author[0000-0002-9632-6106]{Ken J. Shen}
\affiliation{Department of Astronomy and Theoretical Astrophysics Center, University of California, Berkeley, CA 94720, USA}



\begin{abstract}
Tight binary or multiple star systems can interact through mass transfer and follow vastly different evolutionary pathways than single stars. The star TYC 2597-735-1 is a candidate for a recent stellar merger remnant resulting from a coalescence of a low-mass companion with a primary star a few thousand years ago. This violent event is evident in a conical outflow (``Blue Ring Nebula'') emitting in UV light and surrounded by leading shock filaments observed in H$\alpha$ and UV emission. From Chandra data, we report the detection of X-ray emission from the location of TYC 2597-735-1 with a luminosity $\log(L_\mathrm{X}/L_\mathrm{bol})=-5.5$. Together with a previously reported period around 14~days, this indicates ongoing stellar activity and the presence of strong magnetic fields on TYC 2597-735-1. Supported by stellar evolution models of merger remnants, we interpret the inferred stellar magnetic field as dynamo action associated with a newly formed convection zone in the atmosphere of TYC 2597-735-1, though internal shocks at the base of an accretion-powered jet cannot be ruled out.  We speculate that this object will evolve into an FK Com type source, i.e.\ a class of rapidly spinning magnetically active stars for which a merger origin has been proposed but for which no relic accretion or large-scale nebula remains visible.  We also detect likely X-ray emission from two small regions close to the outer shock fronts in the Blue Ring Nebula, which may arise from either inhomogenities in the circumstellar medium or in the mass and velocity distribution in the merger-driven outflow.

\end{abstract}

\keywords{}

\section{Introduction} \label{sec:intro}
Binary and multi-star systems make up over 1/3 of the star systems in the Galaxy \citep{Raghavan+2010}. It is no surprise, then, that closely orbiting binary stars will commonly interact through events that exchange and shed matter from one star to another (e.g., common envelope evolution (CEE)) or end with the complete engulfment of one star by another (e.g., \citealt{Ivanova+2013}). TYC 2597-735-1 is one of only a few examples of a recent ($t \sim$ 2,000 - 5,000 years ago), unobscured (E(B-V) $\sim$ 0.02) stellar merger remnant found to date \citep{2020Natur.587..387H}.
Other examples include \object{HD 233517} \citep{2003ApJ...582.1032J}, \object{BP Psc} \citep{Zuckerman_2008}, \object{TYC 4144 329 2} \citep{Melis_2009}, and FK~Com type stars \citep[see, e.g.,][for a very detailed study of FK Com itself]{2016ApJS..223....5A}.
This makes TYC 2597-735-1 a rare astrophysical laboratory to test models for the physical processes by which stars merge and for the long-term evolution of the stellar remnants left over from such mergers.

One key stellar behavioral shift that may arise as a result of a stellar merger is the generation of a strong magnetic field \citep{Schneider+2016}. Roughly 10\% of intermediate- to high-mass stars ($M_{\star} \gtrsim$ 1.5 $M_{\odot}$) possess strong magnetic fields, whose origins are currently unknown (e.g., \citealt{Donati+2009,Fossati+2015,Grunhut+2017}). \citet{Soker&Tylenda2007} showed that the extended envelope of a merger product may host a large convective region. Paired with a rapid rotation rate that also arises from the merger process, an efficient dynamo may result, giving rise to strong magnetic activity. The presence of magnetic fields has strong implications on the long-term evolution of the stellar merger system. For example, magnetic fields may drive magnetorotational accretion of surrounding material, help slow down rapidly rotating stars via magnetic braking, and power jets (e.g., \citealt{Schneider+2020}). All these mechanisms, over long timescales, have consequences on the final outcome of a stellar merger remnant as it settles to its equilibrium state.

\citet{Soker&Tylenda2007} suggest the strongest magnetic activity tied to a stellar merger's revitalized dynamo may last centuries. Many other recent suspected stellar mergers like red luminous novae are obscured from view by outflows ejected during the merger event itself (e.g., \citealt{Bond+2003, Tylenda+2016}). TYC 2597-735-1 shows evidence that it harbors a persisting, strong magnetic field \citep[e.g., H$\alpha$ emission and variability, radial velocity shifts strongly correlated with the Ca II IRT stellar atmospheric features,][]{2020Natur.587..387H}. One way to test whether its hypothetical magnetic field exists is to look for signs of X-ray emission at TYC 2597-735-1 and compare its luminosity to the remnant's bolometric luminosity.

TYC 2597-735-1 and the last vestiges of its dissolving outflow, the ``Blue Ring Nebula'' (BRN; \citealt{2020Natur.587..387H}), were partially observed by chance with Chandra over a decade ago. Here, we report the findings of this serendipitous \emph{Chandra} observation and present a new \emph{TESS} lightcurve. We explore whether the X-ray emission observed at TYC 2597-735-1 is consistent with dynamo-driven magnetic field activity or other sources near the star. We summarize the stellar properties of TYC 2597-735-1 in Section~\ref{sec:properties}. In Section~\ref{sec:data}, we describe the observations, the source detection methods implemented to retrieve X-ray signal from regions in the BRN and around TYC 2597-735-1 and the X-ray properties of these regions. In Section~\ref{sec:discussion}, we explore different physical mechanisms that may explain the X-ray luminosity and spectral characteristics observed for TYC 2597-735-1 and place physical limits on the properties of the BRN shock region from its X-ray diagnostics. We conclude in Section~\ref{sec:summary} with a summary.

\section{Properties of TYC 2597-735-1}
\label{sec:properties}

\begin{table}
\caption{Present day stellar properties from \citet{2020Natur.587..387H} \label{tab:parameters}}
\begin{tabular}{lcr}
\hline \hline
parameter & symbol & value \\
\hline
distance & $d$ & $1.9 \pm 0.1$ kpc\\
luminosity & $L_\mathrm{bol}$ & $110 \;L_\sun{}$\\
radius & $R_*$ & $11\;R_\sun{}$\\
mass & $M_*$ & $1-2.1\;M_\sun{}$\\
temperature & $T_\mathrm{eff}$ & 5850 K\\
surface gravity & $g$ & 600 cm s$^{-2}$\\
rotational broadening & $v \sin i$ & 6.5 km s$^{-1}$\\
accretion rate & $\dot M_\mathrm{acc}$ & $<1.5 \times 10^{-7}\; M_\sun{}\;\mathrm{yr}^{-1}$\\
RV period & $P_\mathrm{RV}$ & 13.7 days \\
time since merger & & 1000-5000 yr\\
\hline
\end{tabular}
\end{table}

\cite{2020Natur.587..387H} present imaging and spectroscopic information for TYC 2597-735-1 and derive stellar properties, which we summarize in Table~\ref{tab:parameters}. They detect several signs typically associated with stellar activity, including variable H$\alpha$ emission line profiles and radial-velocity (RV)  variations with a period around 13.7~days. \cite{2020Natur.587..387H} argue that line profiles and the parameter space for a potential companion favor an interpretation of the RV data as caused by stellar activity, potentially in the polar regions.

\cite{2020Natur.587..387H} interpret the combination of the bright BRN and the unusual stellar parameters in a stellar merger scenario, where a low-mass ($0.1\;\mathrm{M}_\sun$), short-period companion merged with the primary star a few thousand years ago. In the process of that merger, a circumstellar disk with an inner radius of 0.1~au was formed. Today, material from that disk is accreting onto the central star. The merger also caused the ejection of a conical outflow moving at velocities of around 400~km~s$^{-1}$. Our line-of-sight is close to the axis of the outflow and we see the approaching and receding flows projected onto the sky as the rings of the BRN.
The BRN is seen brightly in the UV as a diffuse ring and, at further distance from TYC 2597-735-1, narrow H$\alpha$ filaments (Figure~\ref{fig:chandraimage}). \cite{2020Natur.587..387H} explain the H$\alpha$ as the forward shock of the outflow, caused by the ejecta driving a shock wave into the circumstellar medium or ISM, while the UV emission should be dominated by molecular H$_2$ emission that is excited by a reverse shock. This reverse shock is caused by the deceleration of the outflow with the circumstellar medium/ISM and now travels backwards through the outflow.

\section{Data analysis} \label{sec:data}
\subsection{Chandra data reduction}
\begin{figure*}
    \centering
    \includegraphics[width=\textwidth]{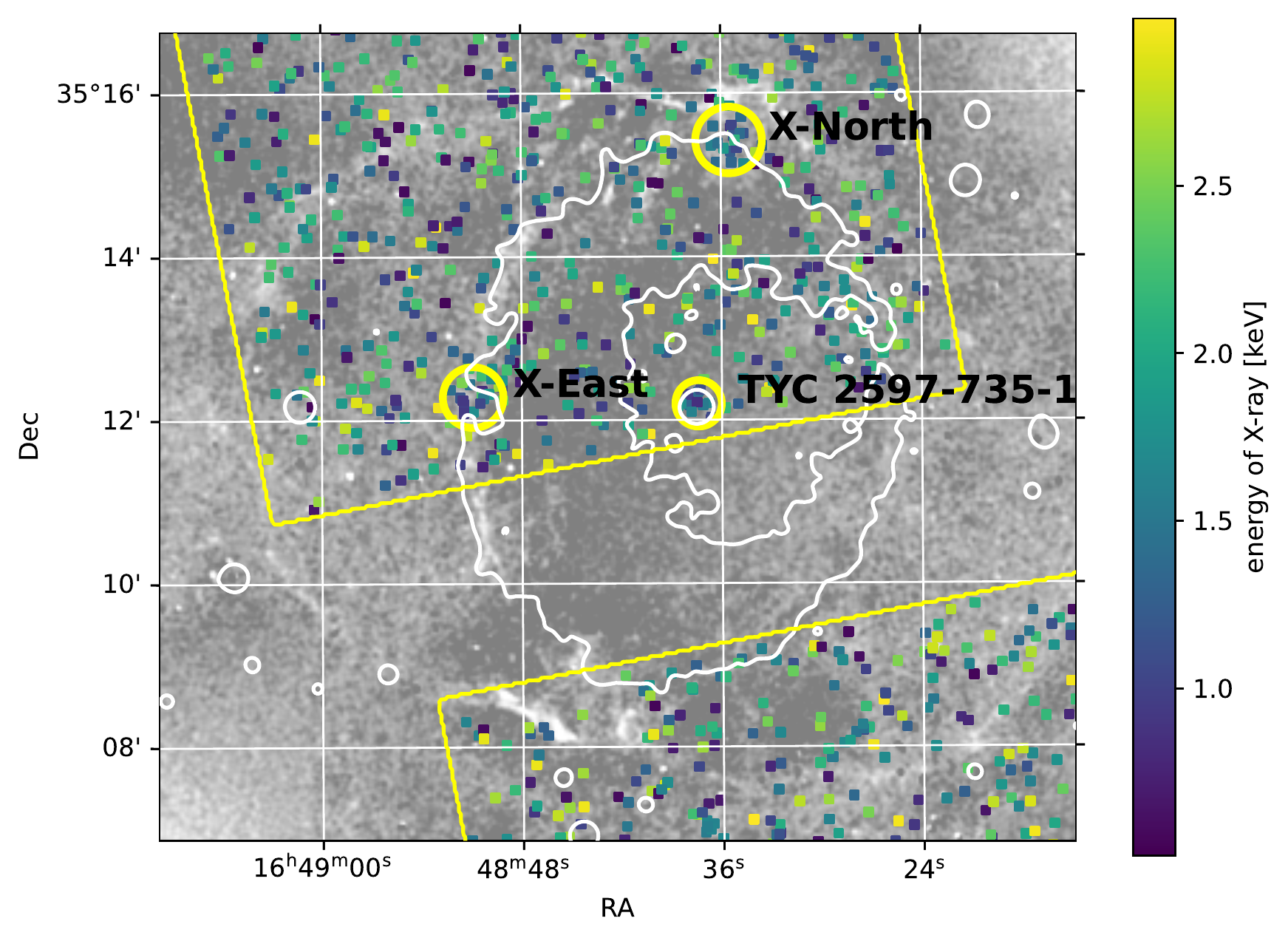}
\caption{X-ray emission around TYC 2597-735-1. The grey-scale image shows H$\alpha$ emission and the white contours outline the UV emission. Data for both is taken from \citet{2020Natur.587..387H}, see there for details. The white contour line is drawn at $6\times 10^{-4}$~counts~s$^{-1}$~pix$^{-1}$ in an image smoothed by a Gaussian filter with $\sigma=4.5$\arcsec{} (3 pixels). This level corresponds to twice the background rate, and is $>10\sigma$ significant in the smoothed image.
    The colored squares mark the position of individual X-ray photons, where the color denotes the energy of each particular photon. Only photons in the energy range given on the color bar are shown. Yellow circles mark the position of the three detected sources and their 90\% PSF size, which we use as the extraction region to extract spectra. The thin yellow lines outline the area covered by Chandra in the observation.}
    \label{fig:chandraimage}
\end{figure*}

TYC 2597-735-1 was observed by Chandra on 2007-11-26 for 8.7~ks.
We retrieved Chandra OBSID 8636 (PI: Murray) from  the archive and reprocessed it with CIAO 4.13.0 \citep{2006SPIE.6270E..1VF} in the VFAINT mode, which reduces the background compared to the default FAINT mode processing. This is a serendipitous observation where TYC 2597-735-1 falls by chance on one of the outer CCDs (CCD S2 on the ACIS-S array). TYC 2597-735-1 is located close to the chip edge and only a fraction of the H$\alpha$ and UV emitting regions around the central star is covered in the observations. Because TYC 2597-735-1 is located off-axis, the PSF is significantly wider than on-axis. Figure~\ref{fig:chandraimage} shows the positions of individual soft X-ray photons overlayed on the H$\alpha$ images and UV emission contours from \citet{2020Natur.587..387H}.

\subsection{X-ray source detection}
\label{sec:xraydetection}
We expect any emission from the outflows to be soft (section~\ref{sec:outflow}), but somewhat absorbed due to the large distance of TYC 2597-735-1. To reduce the high-energy background, which can overwhelm weak, soft sources, we construct a narrow band image in the 0.5-3.0~keV range
and run a source detection with the CIAO \texttt{wavdetect} task, which convolves the input image with a wavelet, taking the size of the PSF at that image location into account. We bin up the image such that it contains $2.6\times10^4$ pixels in the region of interest (the bottom half of the CCD that contains TYC 2597-735-1, marked by the yellow outline in Figure~\ref{fig:chandraimage}) and set the significance threshold such that we expect 0.25 false positive sources in this region. That way, any source detected by \texttt{wavdetect} is likely to be real. We find three sources (table~\ref{tab:src}), all of which coincide with UV/optical emission features. All three sources have similar count rates and similar PSF sizes and thus similar positional uncertainties. We run the CIAO script \texttt{srcflux} to determine the count rate in the presence of background. \texttt{srcflux} performs a fully Baysian computation following \citet{2014ApJ...796...24P} to determine credible intervals.

\begin{table*}
\caption{Detected X-ray sources in the 0.5-3.0~keV band\label{tab:src}}
\begin{tabular}{ccccccc}
\hline \hline
name & RA & DEC & $\sigma_\mathrm{RA}$ & $\sigma_\mathrm{DEC}$ & net counts & 90\% credible interval \\
 &  &  & $\mathrm{{}^{\prime\prime}}$ & $\mathrm{{}^{\prime\prime}}$ & $\mathrm{ct}$ &  \\
\hline
TYC 2597-735-1 & $16^{\mathrm{h}}48^{\mathrm{m}}37^{\mathrm{s}}$ & $35^\circ12{}^\prime12{}^{\prime\prime}$ & 2.9 & 2.0 & 9 & 3-17 \\
X-East & $16^{\mathrm{h}}48^{\mathrm{m}}51^{\mathrm{s}}$ & $35^\circ12{}^\prime17{}^{\prime\prime}$ & 2.7 & 1.8 & 14 & 7-23 \\
X-North & $16^{\mathrm{h}}48^{\mathrm{m}}36^{\mathrm{s}}$ & $35^\circ15{}^\prime25{}^{\prime\prime}$ & 2.2 & 2.3 & 13 & 5-22 \\
\hline
\end{tabular}
\end{table*}

We detect a source at the location of TYC 2597-735-1 with (in our narrow band image) $9$ net counts (90\% credible interval, taking background into account 3-17 counts). The background is known to a high precision because we determine it over a large area. Additionally, the probability that any one source randomly overlaps with the known position of TYC 2597-735-1 within the 90\% PSF radius (0.28 arcmin for an energy of 1.5~keV and at the position of TYC 2597-735-1) is only 0.2\%. Thus, we regard this as a highly reliable detection of X-ray emission from TYC 2597-735-1, albeit with large uncertainties on flux and spectral properties due to the small number of counts.

We detect two other features that are both located just inside the ring of H$\alpha$. One of them is almost due North of TYC 2597-735-1, and the other one to the East (Fig.~\ref{fig:chandraimage}, see table~\ref{tab:src} for source properties).  Neither the GAIA DR2 \citep{2016A&A...595A...1G,2018A&A...616A...1G} nor the 2MASS \citep{2006AJ....131.1163S} catalogs show any point sources coinciding with these features; thus the X-ray emission cannot be caused by a foreground star. 
On the other hand, the vast majority of background objects show hard spectra \citep{2003ApJ...588..696M}, while these sources are soft (see below), making it unlikely they are background objects.
Again, the fact that both features are located at a physically meaningful position between forward shock and the reverse shock of the BRN (seen as H$\alpha$ and the UV emission shown in Fig.~\ref{fig:chandraimage}) argues that these are not background fluctuations, but probably very weak sources that are physically associated with the outflow. However, the number of detected photons is low and only deeper X-ray observations will be able to unambiguously confirm these detections.
All three sources are compatible with point sources. While we expect that the latter two sources are spatially extended, since the H$\alpha$ and the UV emission in the region is also extended, the low count number does not allow us to actually fit the source shape or extent, so we treat them as point sources below.

We also search for extended emission on larger scales that coincides with either the reverse shock where the FUV emission is seen (using a region bounded by the white contours  in Fig.~\ref{fig:chandraimage}) or in the space between forward and reverse shock, i.e.\ in a ring on the outside of the white contour in Fig.~\ref{fig:chandraimage} that contains the sources labelled ``X-North'' and ``X-East''. However, we do not know a priori which level of contour lines might be correlated, so the selection of the region is somewhat arbitrary. We do not see significantly enhanced emission --except for the three point-like sources discussed above-- compared to a background region on the northern half of the chip. To obtain an order-of-magnitude estimate how bright an extended source would have to be for a secure detection, we take the $3\sigma$ uncertainty of the observed background flux (about $10^{-5}$~counts~s$^{-1}$~arcsec$^{-2}$) in the 1-2~keV energy range and convert that the number into an energy flux per surface area in the plane of the sky. We find that emission would have to be stronger than a few times $10^{31}$~erg~s$^{-1}$~pc$^{-2}$ in this band to be detected significantly above the background.

\subsection{X-ray source properties}
\label{sec:xrayspectra}
For all three regions, the detected soft photons are well distributed over the observing time; we do not see strong clustering that would indicate a large flare, but given the low count numbers, variability with factors of a few cannot be excluded.

\begin{figure}
    \plotone{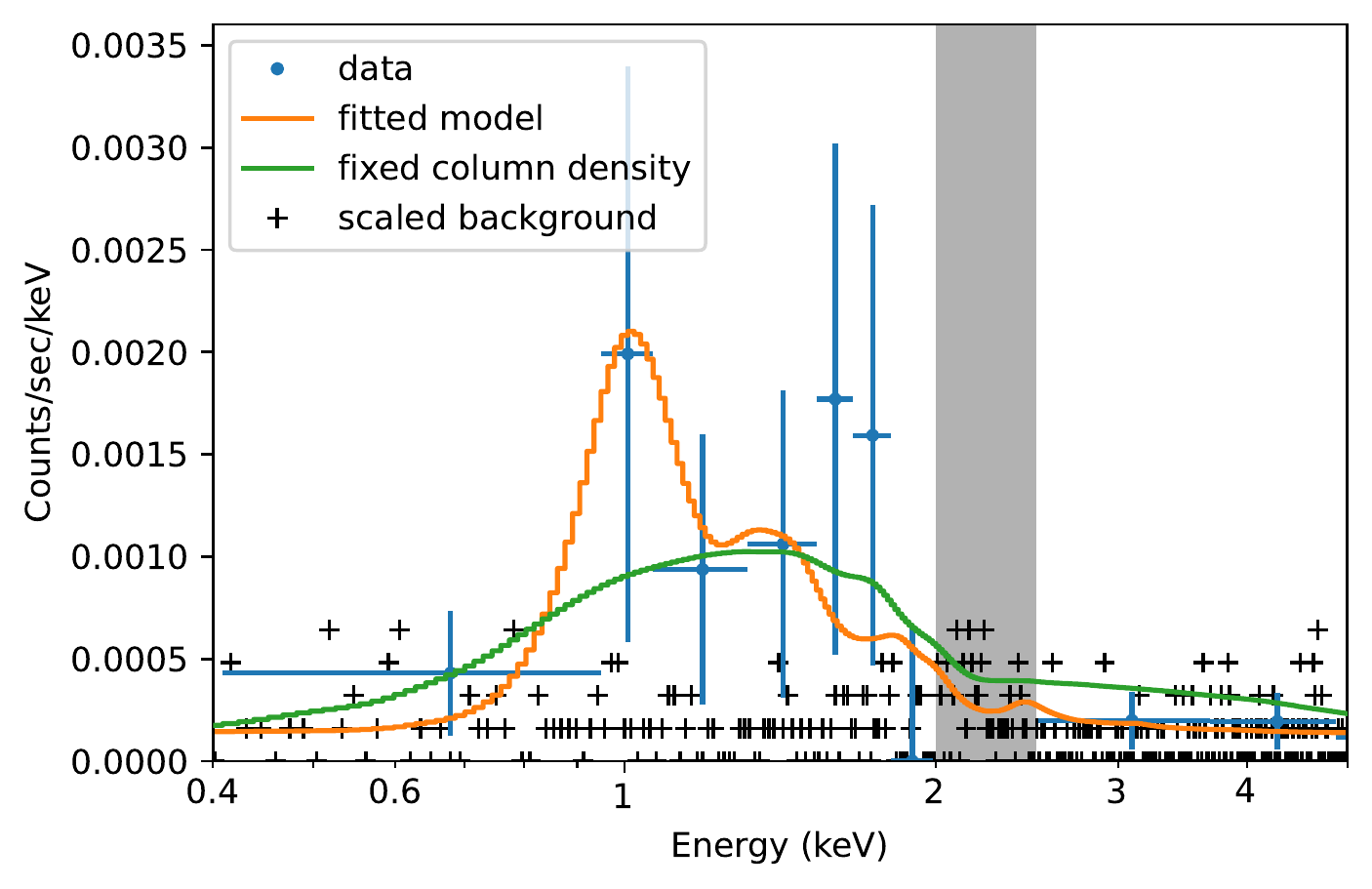}
    \caption{Spectrum of TYC 2597-735-1 binned for display purposes. Error bars shown are calculated as $\sigma_N = \sqrt{N}$ for a bin with $N > 0$ counts and $\sigma_N=1$ for $N=0$. This is just to guide the eye; the fit is performed minimizing the Poisson likelihood on unbinned data. The orange line shows the best fit model with $N_\mathrm{H}$ as a fitted parameter, for the green model $N_\mathrm{H}$ is fixed as described in the text. Black crosses show the background rate, scaled from a much larger area. The background feature that caused us to ignore the region 2.0-2.5~keV (grey) in the fit is also visible. Error bars on the background are omitted for clarity.
    \label{fig:TYC_spec}}
\end{figure}

\begin{figure}
    \plotone{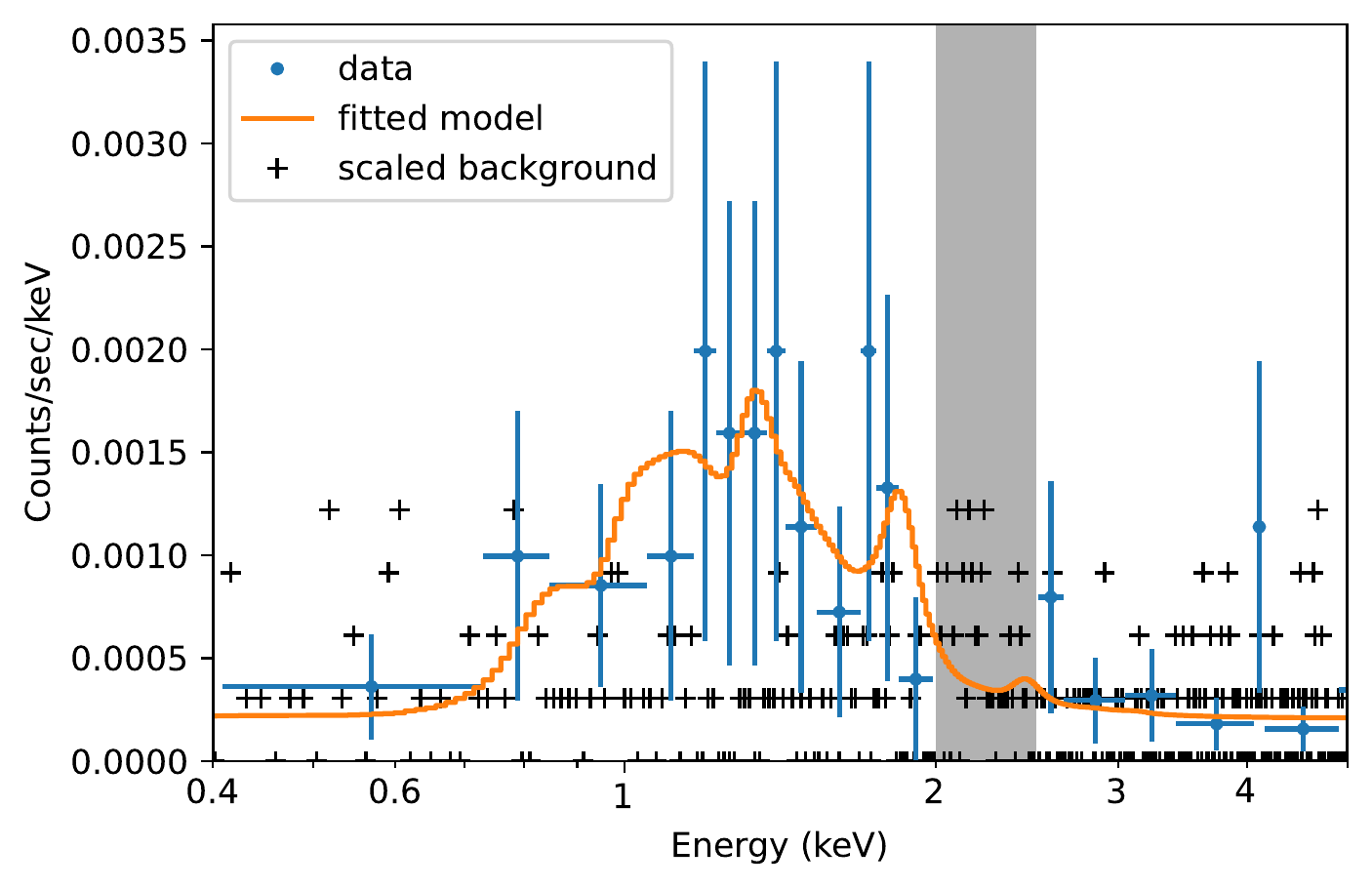}
    \caption{Combined spectrum from the two X-ray detected regions in the outflow. See Fig.~\ref{fig:TYC_spec} for an explanation of symbols.
    \label{fig:combined}}
\end{figure}

We extract spectra for all three sources using circular extraction regions that encompass 90\% of the PSF at 1.5~keV at the location of the source. A background spectrum is extracted from a large region on the same chip located outside of the UV contours shown in Fig~\ref{fig:chandraimage}. Below 7~keV, the background is well described by a flat component, except for a significant feature around 2.2~keV \citep[caused by fluorescence of gold in the instrument,][]{2021arXiv210811234S}. Therefore, we exclude the regions 2.0-2.5~keV and $>7.0$~keV from fitting. We then simultaneously fit the source and background spectra using a Poisson likelihood as implemented in the modified Cash statistic \citep{1979ApJ...228..939C} in Sherpa \citep{2007ASPC..376..543D,doug_burke_2021_4428938}. To fit source emission, we use a model of an absorbed, collisionally-excited plasma \citep[APEC model,][]{2012ApJ...756..128F} and add the flat background model, scaled for the different extraction regions. That way, we avoid subtracting the background and retain Poisson statistics. All fits are done on unbinned data, but the spectra are shown binned for display in Figs.~\ref{fig:TYC_spec} and \ref{fig:combined}.

The spectrum for TYC 2597-735-1 is shown in Fig.~\ref{fig:TYC_spec}. We fit $N_\mathrm{H}=(6\pm5)\times10^{21}$~cm$^{-2}$ and a plasma temperature in the range 0.1-1.5~keV, where the low end of the temperature range requires very large values of $N_\mathrm{H}$ and emission measure. For temperatures around 1~keV, common for stellar sources, the unabsorbed flux is of order $3.6\times10^{-14}$~erg~s$^{-1}$~cm$^{-2}$, which corresponds to $1.3\times10^{30}$~erg~s$^{-1}$ with uncertainties of a factor of a few.

The total galactic absorption in this line of sight is about $N_\mathrm{H}=2\times10^{21}$~cm$^{-2}$ \citep{1990ARA&A..28..215D}. Fitting just the temperature, but fixing the absorbing column density leads to a model with similar spectral shape (green line in Fig.~\ref{fig:TYC_spec}), but a preferred higher temperature $kT=4_{-3}^{+\infty}$~keV.
For low count numbers, there is a fundamental degeneracy between plasma temperature and absorbing column density: A cool, but bright plasma with a high $N_\mathrm{H}$ produces a spectrum very similar to the spectrum emitted from a hotter, but less absorbed plasma.
Assuming the hotter spectrum, a weaker source ($2\times10^{-14}$~erg~s$^{-1}$~cm$^{-2}$, which corresponds to $8\times 10^{29}$~erg~s$^{-1}$) could be sufficient to explain the observed photons; on the other hand, there could still be a second, cooler, component that is hidden by the $N_\mathrm{H}$, thus the flux given is a lower limit only. \cite{2020Natur.587..387H} found essentially no reddening towards TYC 2597-735-1, but depending on the position of the X-ray emitting gas it is possible that accretion columns or the disk are in our line-of-sight and that the $N_\mathrm{H}$ observed in X-rays is larger than the reddening of the optical spectra.

We combined the spectra for X-East and X-North because the count number is very low and X-ray properties (see the color of the detected photons in Fig.~\ref{fig:chandraimage}) and location (between H$\alpha$ ring and UV emission) are similar.
We fit $N_\mathrm{H}=(1.3_{-0.3}^{+0.5})\times10^{22}$~cm$^{-2}$ and a plasma temperature of $0.5_{-0.3}^{+\infty}$~keV (Fig.~\ref{fig:combined}). For temperatures around 0.5~keV, which we will argue are physically reasonable in Sect~\ref{sec:outflow}, the unabsorbed flux is of order $4.4 \times 10^{-13}$~erg~s$^{-1}$~cm$^{-2}$, which corresponds to $1.5\times10^{31}$~erg~s$^{-1}$ with uncertainties of a factor a few. As in the case for TYC 2597-735-1, a hotter source with a lower flux seen through a lower absorbing column density could also generate the observed spectrum.

\subsection{Optical photometry}
\label{sec:TESS}
The Transiting Exoplanet Survey Satellite (TESS; \citealt{Ricker2015}) observed TYC 2597-735-1 in its sector 25 and camera 1 from 2020-May-13 until 2020-Jun-08.
TESS is an optical photometry mission with four 24x24 degree cameras operating in the 600-1000\,nm range. The spacecraft scans through the entire sky in sectors of $\sim28$\,day duration, whereby each sector is divided into two satellite orbits of $\sim14$\,day duration.
We identify TYC 2597-735-1 as TIC 284856863 in the TESS Input Catalog version 8 \citep{Stassun2019} and find that it fell into TESS's full-frame images with 30 min cadence.
We retrieve the mission's official MIT quick-look-pipeline (QLP) light curve from the High-Level Science Products (HLSP) via the Mikulski Archive for Space Telescopes (MAST) archive.
We also estimate the background flux from the target pixel files (TPFs) via the tools \texttt{tesscut} \citep{Brasseur2019} and \texttt{eleanor} \citep{Feinstein2019}.

\begin{figure*}
    \centering
    \includegraphics[width=0.8\textwidth]{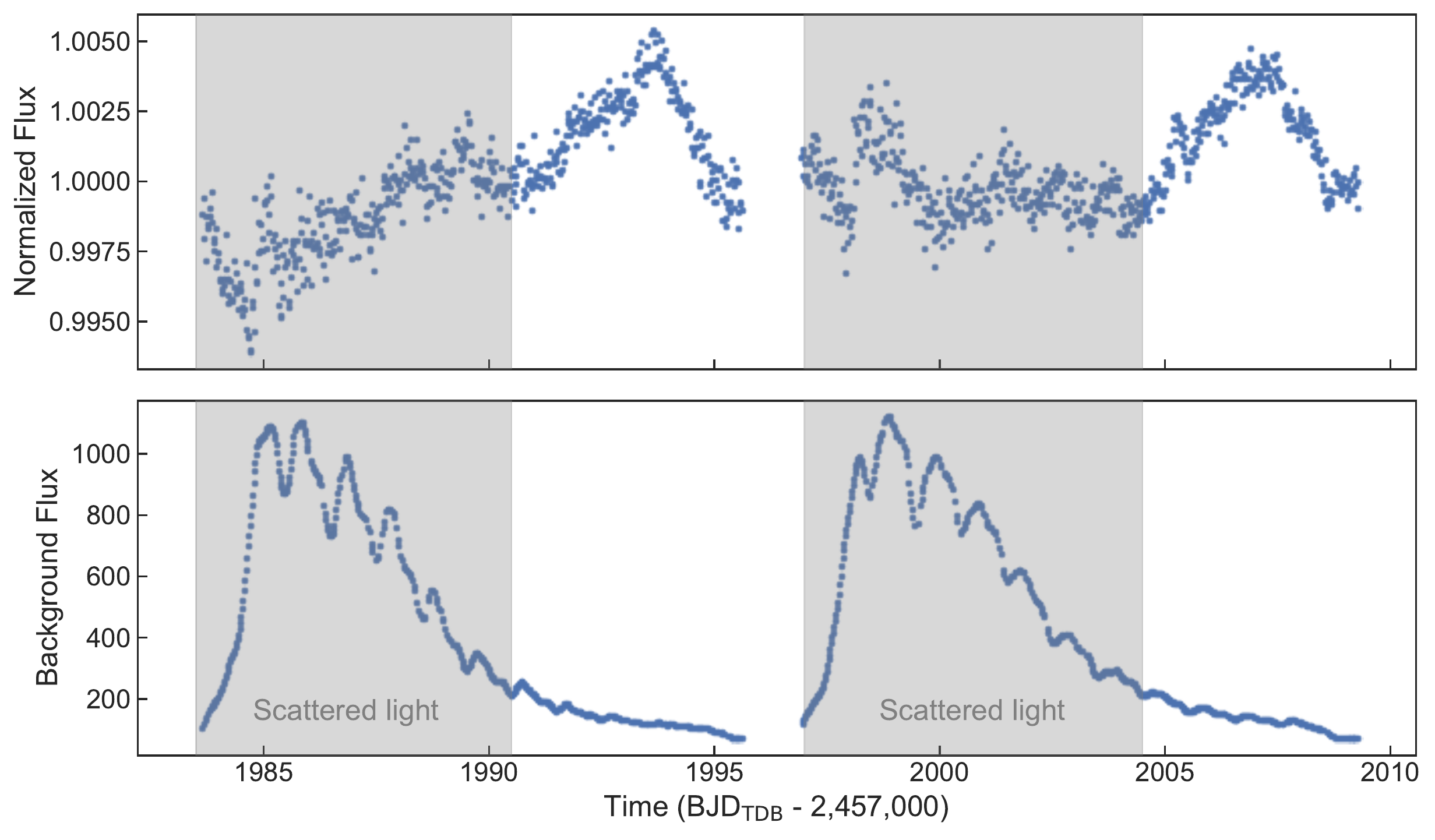}
    \caption{TESS light curve and background flux for TYC 2597-735-1. The observations taken in sector 25 with camera 1 range from 2020-May-13 until 2020-Jun-08. The light curve has been corrected for background effects and displays significant variability. While the start of each TESS orbit suffer from systematics caused by Earth's scattered light (grey-shaded areas), the end of each orbit shows brightness increases that could be of astrophysical nature.
    \label{fig:TESS_light_curve}}
\end{figure*}

Inspecting the background- and systematics-corrected light curve and corresponding background flux, we find significant variability (Fig.~\ref{fig:TESS_light_curve}).
This could be due to either intrinsic stellar variability or instrument systematics, or a combination of both.
On the one hand, the radial velocity signal from TYC 2597-735-1 also showed a period of 13.7 days \citep{2020Natur.587..387H}. This would match the period of the two major brightness increases towards the ends of each TESS orbit (TJD 1,990-1,995 and 2,005-2,010; where TJD = $\mathrm{BJD_{TDB}}$ - 2,457,000).

On the other hand, the TESS data release notes\footnote{\url{https://tasoc.dk/docs/release_notes/tess_sector_25_drn36_v02.pdf}} state that Earth's scattered light particularly affected this sector and camera. These notes, along with our inspection of the local background flux (Fig.~\ref{fig:TESS_light_curve}) and the light curves of neighboring stars of similar brightness (not shown here), suggest that these systematics affected mainly the start of each orbit. This means the end of each orbit should remain reliable.

However, the similarity of the expected stellar rotation period and the TESS orbit duration (both $\sim14$\,days) make it difficult to disentangle the true origin of the variability. TESS will re-observe TYC 2597-735-1 in sectors 51 and 52 (from 2022-Apr-22 until 2022-Jun-13). This extended observing window, the possibility of increased observation cadence (2\,min or even 20\,s cadence), and the improved visibility will help to better understand TYC 2597-735-1's behavior at optical wavelengths.

\cite{2020Natur.587..387H} noticed an RMS scatter around 0.008~mag --significantly larger than the formal uncertainties-- 
between observations taken with different telescopes in 2015 and in Nov 2019.
They attributed this to systematic calibration uncertainties between different telescope systems. However, given the variability in the TESS lightcurve on a similar level, this might instead be a signature of stellar rotation leading to B-band variability on time scales of a few days.

\section{Discussion}  \label{sec:discussion}
In stellar sources, X-rays mostly originate from optically thin, collisionally excited plasma with temperatures of a few MK. The plasma can be either heated by magnetic reconnection as in the solar corona, or in shocks (accretion shocks or shocks in winds or jets). Unless the density is very low ($<10^5$~cm$^{-3}$ or so), we can assume that the plasma is in an equilibrium ionization state \citep{2007A&A...466.1111G}.
Before we discuss possible mechanisms to generate X-rays for TYC 2597-735-1 in Sect.~\ref{sec:discussTYC} and for the resolved outflow in Sect.~\ref{sec:outflow}, we summarize the density and temperatures in strong hydrodynamical shocks, since several possible scenarios involve shock heated plasma; for a more detailed derivation of the following equations see e.g.\ \citet{2007A&A...466.1111G} and \citet{1967pswh.book.....Z}.

Assume material of velocity $v_{\rm ej}$ is moving into a stationary external medium.  The velocity of the forward shock in the lab frame $v_{\rm fsh}$ is given by the shock jump conditions (for adiabatic index $\gamma = 5/3$),
\be
v_{\rm fsh} = \frac{4}{3}v_{\rm psh}, \label{eqn:vsh1}
\ee
where $v_{\rm psh}$ is the velocity of the post-shock gas.  The reverse shock velocity (also in the lab frame) is given by
\be
v_{\rm rsh} \simeq \frac{4}{3}v_{\rm psh} - \frac{v_{\rm ej}}{3}
\label{eqn:vsh2}
\ee
From the jump conditions, the shocks will heat gas to a temperature
\be
kT_{\rm sh} \simeq \frac{3}{16}\mu m_p v_{\rm sh}^{2} \approx 0.3\,{\rm keV}\left(\frac{v_{\rm sh}}{500\,{\rm km/s}}\right)^{2},
\label{eqn:Tshock}
\ee
where we have taken $\mu \simeq 0.62$ for fully ionized solar composition gas and $v_{\rm sh}$ is the shock velocity (i.e., $v_{\rm fsh}$ or $v_{\rm rsh}$) relative to the upstream rest frame. $m_p$ is the mass of a proton.

\subsection{TYC 2597-735-1}
\label{sec:discussTYC}
A variety of physical mechanisms could in principle be responsible for the X-ray emission observed at TYC 2597-735-1. The point of interest is that TYC 2597-735-1 is strongly suspected of undergoing a merger event with a smaller, $\sim$0.1 M$_{\odot}$ companion $\lesssim$5,000 years ago \citep{2020Natur.587..387H}. Such a collision and subsequent engulfment of a stellar companion would disrupt the equilibrium state of the surviving star, potentially leading to enlarged surface convective zone regions (e.g., \citealt{1996ApJ...468..797L}), accretion flows from surrounding circumstellar material that continue to disturb the star's outermost surface regions (e.g., \citealt{2016MNRAS.461.2527P}), and sustained outflows that continue to eject material from the system (e.g., \citealt{Zuckerman_2008}). Here, we explore different mechanisms that may be responsible for the X-ray emission observed at TYC 2597-735-1.

\subsubsection{X-rays from Accretion Shocks onto TYC 2597-735-1}

Current observations suggest TYC 2597-735-1 harbors a substantial circumstellar disk, which may have an inner rim as close as $\sim$0.1 AU \citep{2020Natur.587..387H}. Much like planet-forming accretion disks, complex interactions between these disks and stars via magnetospheric accretion can lead to matter falling onto the star, potentially generating X-rays at free-fall shock regions \citep[e.g.][]{2007A&A...466.1111G}.

With the stellar parameters from Table~\ref{tab:parameters} we calculate the free fall velocity of material onto TYC 2597-735-1 as
$v_{\rm ff} \lesssim$ 260 km/s. At this free fall velocity, the temperature of the shocked material according to Eqn~\ref{eqn:Tshock} for $v_{\rm sh} \approx v_{\rm ff}$ is $\lesssim8\times 10^5$~K $-$ too low to explain the observed X-ray flux.
Instead, the accretion shock should peak in luminosity in the EUV ($\sim$180 - 280 {\AA}) where the total shock luminosity is expected to be a small fraction of $L_{\star}$:
\begin{equation}
    L_{\rm sh} = \frac{1}{2} \dot{M}_{\rm acc} v_{\rm ff}^{2} \approx 3.5 \times 10^{-3} L_{\odot}\left(\frac{\dot{M}_{\rm acc}}{10^{-7}M_{\odot}{\rm yr^{-1}}}\right),
\end{equation}
or $L_{\rm sh}/L_{\rm bol} <$ 3$\times$10$^{-5}$.

\subsubsection{Shocks at the base of the outflow}
\label{sect:baseoutflow}
In the young star DG~Tau, X-rays have been observed at the base of the outflow \citep{2008A&A...488L..13S} possibly related to the launching or collimation \citep[e.g.][]{2014ApJ...795...51G,2018A&A...615A.124U}. In DG~Tau, the soft X-ray emission is offset by about 40~au, but given the source location offset from \emph{Chandra}'s optical axis, the distance of TYC 2597-735-1, and the viewing geometry where the outflow axis is close to the line-of-sight, we do not expect an obervable positional offset here.

If we assume that a fraction $\eta_{\rm j} \sim 10\%$ of the accreted matter is launched into a bipolar collimated outflow (``jet'') with an outflow velocity $v_{\rm j}$ some multiple $\sim 2$ of the surface escape speed of TYC 2597-735-1, $\approx (2GM_{\star}/R_{\star})^{1/2} \approx 230$ km s$^{-1}$, then the present-day jet power is given by
\be
L_{\rm j} \approx \frac{1}{2}\eta_{\rm j}\dot{M}_{\star}v_{\rm j}^{2} \approx 5\times 10^{32}\,{\rm erg\,s^{-1}}\,\left(\frac{\eta_{\rm j}}{0.1}\right)\left(\frac{\dot{M}_{\rm acc}}{10^{-7}M_{\odot}\,\rm yr^{-1}}\right)\left(\frac{v_{\rm j}}{400\,{\rm km\,s^{-1}}}\right)^{2},
\label{eqn:jetpower}
\ee
This is sufficient to explain the X-ray point source overlapping TYC 2597-735-1, $L_{\rm X} \sim 10^{31}$ erg s$^{-1}$, if the entire outflow reaches internal shock speeds of order 500~km~s$^{-1}$ (Eqn.~\ref{eqn:Tshock}).
In young stars, outflows often have different velocity components, where only the fastest and innermost components shock to produce X-rays, while most of the mass in seen in slower components. For example, in DG~Tau, the measured velocity in the UV is similar to the outflow velocity of the BRN and only a small fraction of the mass loss is fast enough to generate X-rays. However, such a scenario would require an unreasonably large mass-loss, since Eqn.~\ref{eqn:jetpower} already requires that 10\% of the total accretion rate shocks to produce X-rays. Alternatively, a small fraction of the outflow could be accelerated to several times the escape speed to provide a small, but hot and luminous X-ray source in contrast to what is observed in other systems with X-rays at the jet base such as DG~Tau \citep{2009A&A...493..579G}.

\subsubsection{Activity powered by a stellar dynamo}
\label{sec:dynamo}
The Rossby number, or the ratio of the rotation period $P_{\rm rot}$ to the convective overturn time $\tau_{\rm conv}$, i.e.
\begin{equation}
{\rm Ro} \sim \frac{P_{\rm rot}}{\tau_{\rm conv}}, \label{eqn:ro}
\end{equation}
exhibits notable correlations with main sequence stellar magnetic activity and is well established in the frame of stellar dynamo models (e.g., \citealt{Brandenburg+1998}).
It can also be used to compare the level of magnetic activity for pre-main sequence, main sequence, and giant stars \citep{Preibisch+2005, Pizzolato+2003, Gondoin+2005}.

\citet{2020Natur.587..387H} calculate a grid of \texttt{MESA} models to simultaneously track the temporal behavior of a recent stellar merger to match its properties to observed characteristics of TYC 2597-735-1 and estimate the original binary star properties (in particular, M$_{\star}$) that best matches the anticipated age (from the expansion rate of the BRN) and observed properties of TYC 2597-735-1 today. We use the model calculation from \citet{2020Natur.587..387H} that is closest to the observed properties of TYC 2597-735-1, but we note that the model is not tuned to the exact parameters of TYC 2597-735-1. We extract the properties of possible convection zones that may have arisen in TYC 2597-735-1 post-merger with a 0.1 M$_{\odot}$ companion. The primary star in the \texttt{MESA} evolutionary model is set to a pre-merger mass M$_{\star}$ of 2.17 M$_{\odot}$ and a pre-merger radius R$_{\star}$ of 5 R$_{\odot}$. The models suggest the primary (surviving) star may have two prominent convection zones that persist $>$5,000 years after the stellar merger: a deep convection zone ($\sim$0.5 R$_{\odot}$) and a surface convection zone.
As the dynamo near the stellar surface is the one that likely controls magnetic activity, we focus on the properties of the surface convection zone of our modeled star.

\begin{table}

\caption{Surface Convection Zone Properties of TYC 2597-735-1
	\label{tab:mesa}}
\begin{tabular}{r c c c c | c l}
\hline\hline
\multicolumn{5}{c|}{Best-fit \texttt{MESA} Evolutionary Model} & \multicolumn{2}{c}{derived value} \\
Time & $R_{\star}$ & $r_{\rm conv}$ & $v_{\rm conv, max}$ & $\tau_{\rm conv} = \frac{r_{\rm conv}}{v_{\rm conv, max}}$ & Ro & ${\rm C_X}$ \\
(years)    &    ($R_{\odot}$)   & (10$^{10}$ cm)           & (10$^{6}$ cm/s)        & (days) \\
\hline
	100  & 14.9 & 0.774 & 1.33 & 0.0673 & 200 & 0.1  \\
	200  & 12.9 & 0.534 & 1.23 & 0.0503 & 270 & 0.2  \\
	500  & 11.0 & 0.581 & 1.05 & 0.0642 & 210 & 0.1  \\
	1000 & 9.9  & 2.12 & 0.785 & 0.312  & 44 & 0.006  \\
	2000 & 8.8  & 4.24 & 0.559 & 0.877  & 16  & $7\times10^{-4}$\\
	5000 & 7.4  & 7.63 & 0.341 & 2.589  & 5   & $8\times10^{-5}$ \\
\hline
\end{tabular}
\tablecomments{See section~\ref{sec:dynamo} for a detailed explanation of columns. The last two columns are not part of the \texttt{MESA} model. They have been derived using eqn.~\ref{eqn:ro} and \ref{eqn:cx} using the period and $L_\mathrm{bol}$ from table~\ref{tab:parameters} and the X-ray flux from section~\ref{sec:xrayspectra}.}
\end{table}

Table~\ref{tab:mesa} presents the primary star's stellar radius ($R_{\star}$), total height of the surface convection zone ($r_{conv}$), the maximum convective velocity within the convective region ($v_{conv,max}$), and the corresponding overturn time ($\tau_{\rm conv}$) at different snapshots in time since $t=0$ of the complete stellar merger from the \texttt{MESA} model that comes closest to the observed properties of TYC 2597-735-1 \citep{2020Natur.587..387H}. There are some discrepancies between model values and observed quantities. In particular, the current observed radius of $R_*=11R_{\odot}$ is predicted by the model at $t=500$~yr (table~\ref{tab:mesa}), while the expansion velocity of the BRN points to a time since merger of 2000-5000~yr (table~\ref{tab:parameters}). Given uncertainties on the observed present-day stellar parameters and on the assumptions in the model itself \citep[see supplementary information of][]{2020Natur.587..387H}, as well as the fact that the model is not meant to be an exact match to TYC 2597-735-1, but rather the closest match in a grid of models, those differences are not surprising. The main goal of table~\ref{tab:mesa} is to present an estimate of $\tau_{\rm conv}$ to allow us to calculate
the Rossby number (${\rm Ro}$) at each timestep in our model assuming the present-day rotation period. The measured $L_X$ is uncertain by factors of a few (section~\ref{sec:xrayspectra}), and dominates over the uncertainties from the model in table~\ref{tab:mesa}.
The merger itself inflates the stellar radius, which means the rotation period at earlier times would have to be longer than the 13.75 day rotation observed by \citet{2020Natur.587..387H} and TESS (Section~\ref{sec:TESS}) to conserve angular momentum. Even then, Ro is very large ($>$20) for t$\lesssim$2,000 years.  \citet{Pizzolato+2003} show that magnetic activity in main sequence stars ceases for ${\rm Ro} \gtrsim 10$, suggesting that for much of the time close to the merging of the two stars, TYC 2597-735-1 may not have significant magnetic activity.
At later times ($t\gtrsim$2,000 years), ${\rm Ro}$ is within the regime where magnetic activity might be present.

The dynamo generated X-ray emission exhibits an empirical luminosity of the form  (e.g., \citealt{Soker&Tylenda2007})
\begin{equation}
\frac{L_{\rm X}}{L_{\rm bol}} = C_{\rm x}{\rm Ro}^{-2}, \label{eqn:cx}
\end{equation}
where the constant $C_{\rm x}$ varies with the stellar system. We present the $C_{\rm x}$ that matches the observed X-ray flux for each timestep in the \texttt{MESA} model in Table~\ref{tab:mesa}. \citet{Soker&Tylenda2007} note that $C_{\rm x}$ is valid for 0.15 $\lesssim$ ${\rm Ro} \lesssim$ 10, where $L_{\rm X} / L_{\rm bol}$ saturates at values $\sim$10$^{-3}$ (${\rm Ro} \lesssim$ 0.15). In the 2,000-5,000~yr range of our models, $C_{\rm x}$ ranges from 10$^{-3} - 6\times$10$^{-5}$. The lower end is in line with typical main sequence-to-early subgiant magnetic activity levels ($C_{\rm x} \sim$ 10$^{-5}-10^{-6}$; \citealt{Pizzolato+2003, Gondoin+2005}).


\subsubsection{Comparison to other recent merger candidates}
\label{sect:comparison}
\begin{figure}
    \plotone{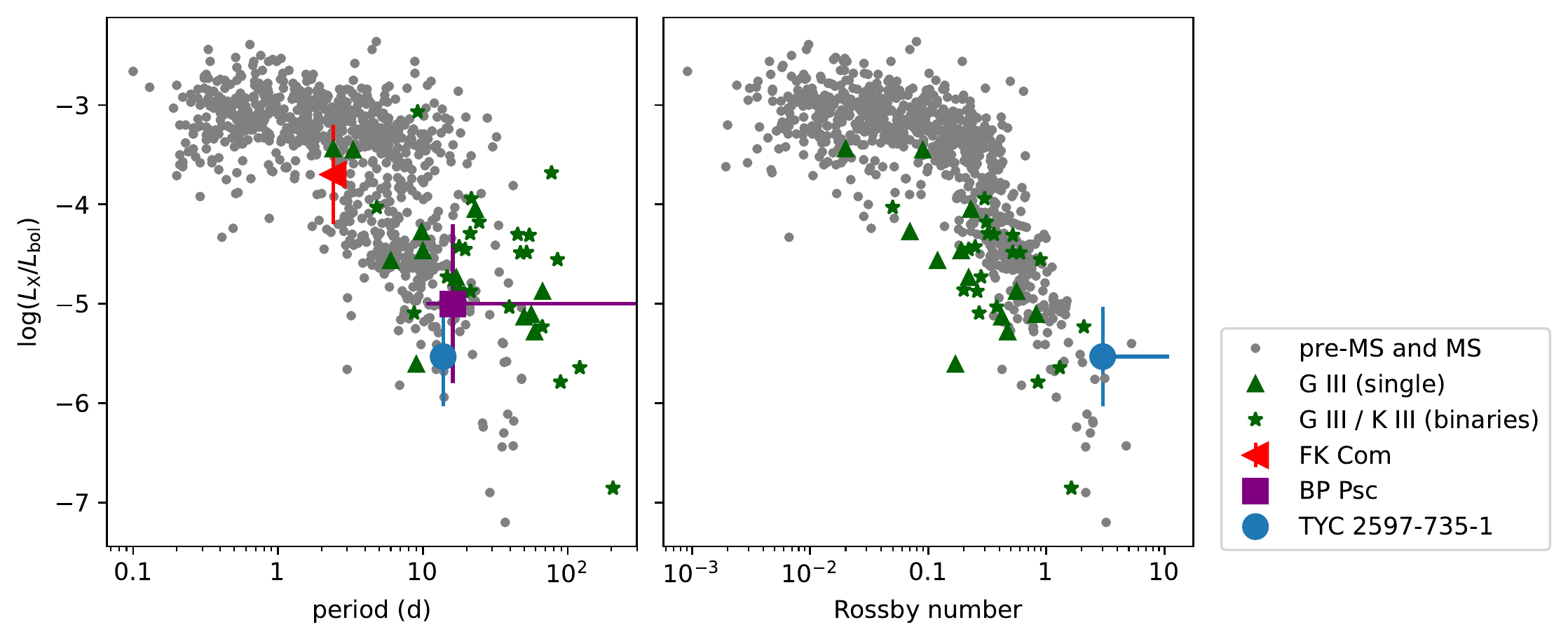}
    \caption{X-ray activity level for TYC 2597-735-1 compared to some other dwarfs and giants as a function of rotation period (left) and Rossby number (right), see text for a discussion of uncertainties adopted for TYC 2597-735-1 and references for the samples shown. Error bars are omitted for clarity for samples with more than five stars.
    \label{fig:lxbol}}
\end{figure}

Figure~\ref{fig:lxbol} compares the activity level in the form $\log(L_\mathrm{X}/L_\mathrm{bol})$ to a sample of dwarf stars from \citet{Wright+2011}. This sample includes pre-main sequence stars, young active stars, and older main-sequence stars. The activity level increases with faster rotation and saturates around $\log(L_\mathrm{X}/L_\mathrm{bol})=-3$. Pre-main sequence stars and young stars are generally found in the saturated regime, while the activity, and thus the X-ray luminosity, declines over time on the main-sequence as stars loose angular momentum and spin down.  A sample of single giant stars is taken from \citet{Gondoin+2005} and giants that are part of binary systems from \citet{Gondoin+2007}. Figure~\ref{fig:lxbol} (left) shows the relation to the rotation period, which is directly observed. Comparing $\log(L_\mathrm{X}/L_\mathrm{bol})$ to the Rossby number requires knowledge of the internal structure of the star to estimate the convective turnover time. \citet{Wright+2011} fit an empirical relation to their sample (gray dots in  Figure~\ref{fig:lxbol}), while \citet{Gondoin+2005} and \citet{Gondoin+2007} rely on stellar structure models from \citet{Gunn1998}, which are only available up to $2.2\;M_\sun$.

We use the period detected in the RV curves which is consistent with our TESS lightcurve to place TYC 2597-735-1 in the figure; for the Rossby number, we plot the range indicated in table~\ref{tab:mesa} for 2000-5000~yr post-merger. For comparison, using the tracks shown in \citet{Gunn1998} would give a Rossby number between 1.4 and 40 for the mass range of $1-2.1\;M_\sun{}$ (Table~\ref{tab:parameters}).

For comparison, we include \object{FK Com}, the prototype of the FK Comae Berenices (FK Com) class of variable stars. FK~Com shows a periodic lightcurve with a period of $P=2.4$~d \citep{1993A&A...278..449J}, and a number of other activity indicators, such as hot Si~{\sc iv} and chromospheric Ca~{\sc ii} and C~{\sc ii} emission lines, as well as persistent and flaring X-ray activity \citep{2016ApJS..223....5A}. Similar to our target, FK~Com stars are also thought to be the result of recent close-binary mergers, which caused the remaining star to spin up and thus triggers a dynamo and coronal activity. Unlike our target TYC 2597-735-1, FK~Com stars are point sources and no BRN is seen. On the other hand, TYC 2597-735-1's BRN will fade with time as the shocks weaken and the density of the outflow decreases because it expands while the central star TYC 2597-735-1 itself, which is currently overluminous because of its large size, should contract towards a stable equilibrium and conservation of angular momentum will cause it to spin up. In other words, TYC 2597-735-1 might become an FK Com star in the future.

\object{BP Psc} is another giant suspected to be the result of a recent merger. Unlike the FK~Com stars, BP~Psc does have circumstellar material, including a disk and a bright jet \citep{Zuckerman_2008}.
\citet{2010ApJ...719L..65K} observed $\log(L_\mathrm{X}/L_\mathrm{bol})$ between -5.8 and -4.2; the absorbing column density is very high, leaving the amount of soft X-ray emission unconstrained. Unfortunately, the period of BP~Psc is not known, but we can derive some constraints. For this, we assume that BP~Psc is a post main-sequence object and not, as has sometimes been suggested, a young star \citep[see][for a discussion of the problems with young star scenarios]{Zuckerman_2008}. Interpreting its spectrum in the post main-sequence scenario, \citet{Zuckerman_2008} derive $M_*=1.8M_\sun$ and $\log g=2.5\pm0.3$, which leads to a radius around $12R_\sun$.
Assuming that the stellar rotation axis is aligned with the symmetry axis of the disk ($i=75\pm10$ degree), we can convert the upper limit on the rotational broadening of $v \sin i=38$~km~s$^{-1}$ from \citet{Zuckerman_2008} into a lower limit of the period of 19~d and place BP~Psc in Fig.~\ref{fig:lxbol}.

Figure~\ref{fig:lxbol} shows that all these different classes of objects behave remarkably similar: Despite some scatter, stars of different sizes and with different evolutionary histories show a clear increase in $\log(L_\mathrm{X}/L_\mathrm{bol})$ with decreasing period. The (single or binary) field giants tend to be a little to the right (in the left panel) of the MS stars, i.e.\ they reach a similar level of $\log(L_\mathrm{X}/L_\mathrm{bol})$ for a longer rotation period than MS stars; on the other hand, they are to the left of MS stars in Rossby number (right panel). Our recent merger candidates (FK Com, TYC 2597-735-1, and BP Psc) are compatible with MS stars and field giants.
However, TYC 2597-735-1 where we have stellar structure simulations to estimate the Rossby number, shows a larger Rossby number for a given $\log(L_\mathrm{X}/L_\mathrm{bol})$.

\subsubsection{What is the source of X-rays from TYC 2597 735 1?}
The discussion above shows that the X-ray emission could be explained by either a shock at the base of a fast outflow component or coronal activity. In either case, magnetic fields need to be present on TYC 2597-735-1 to either accelerate the outflow or to power the corona, so the detection of X-ray emission provides circumstantial evidence of magnetic fields either way. Given that similar $L_\mathrm{X}/L_\mathrm{bol}$ values are observed for comparable sources without accretion disks and outflows (Fig.~\ref{fig:lxbol}) and an outflow shock requires either to shock the entire outflow or an outflow component much faster than the escape speed (section~\ref{sect:baseoutflow}), it seems more likely that emission can be attributed to coronal activity. Deeper observations might be able to distinguish these possibilities: Coronal activity should be time-variable on scales of hours or days, while emission from a jet collimation shock is generated on larger spatial scales and thus varies more slowly. Also, longer observations might be able to resolve the spectrum into more temperature components; a corona might show hot plasma out of reach of a jet shock.

\subsection{Large scale outflow}
\label{sec:outflow}
We now look at what limits on density and mass flux can be placed based on the detection of X-East and X-North (Sect.~\ref{sec:xEN}) and the upper limit on X-ray emission from the large region between forward and reverse shock (Sect.~\ref{sec:ISM}).


We assume that the BRN ejecta of density $n_{\rm BRN}$ is expanding at $v_{\rm ej} \simeq 400$ km/s into a stationary external medium of particle density $n_{\rm ism}$. Given the bright UV emission, we expect that the shock has already swept up considerable gas from the external medium and begun to decelerate, then $v_{\rm fsh} \ll v_{\rm ej}$ and $v_{\rm rsh} \simeq -v_{\rm ej}/3$, using Eqns~\ref{eqn:vsh1} and \ref{eqn:vsh2}.  Here, $v_{\rm sh}$ in Eqn.~\ref{eqn:Tshock} is the shock velocity relative to the upstream rest frame (i.e. $v_{\rm sh} \simeq v_{\rm fsh}$ in the case of the forward shock with stationary upstream, but $v_{\rm sh} \simeq v_{\rm ej}-v_{\rm rsh}$ in the reverse shock).  From the above, we expect the reverse shock will be strongest, with $v_{\rm sh} \simeq (4/3)v_{\rm ej} \approx 500$ km/s.

This argument makes it reasonable to interpret the X-ray spectra of X-East and X-North as a plasma with a temperature around 0.3-0.5~keV, as we have done in Section~\ref{sec:xrayspectra}.

\subsubsection{Sources X-East and X-North}
\label{sec:xEN}
The sources X-East and X-North are compatible with point sources. Given the size of the PSF, we use 1~arcsec ($d_\mathrm{sh}=1900$~au at the distance of TYC 2597-735-1) as an upper limit of the source size and 500~km~s$^{-1}$ as the shock velocity for the ejecta, compatible with the observed X-ray spectrum.

\citet{2002ApJ...576L.149R} provide a
semi-analytical formula for the cooling length
$d_{\mathrm{cool}}$ of gas
heated in fast shocks ($\Delta v \ga 200$~km~s$^{-1}$) caused by ramming into an obstacle,
which we can rewrite as:
\begin{equation}
d_{\mathrm{cool}} \approx 20.9 \mathrm{ AU}
    \left(\frac{10^5\mathrm{ cm}^{-3}}{n_0}\right)
    \left(\frac{v_{\mathrm{sh}}}{500\textnormal{ km s}^{-1}}\right)^{4.5},
\label{raga}
\end{equation}
where $n_0$ is the pre-shock particle number density, equal to a
quarter of the post-shock number density $n$ in the strong shock
approximation.

Since $d_{\mathrm{cool}}$ is much smaller than the size scale of the outflow we can neglect spherical expansion over the thickness of the shock.
To fit such a shock into 1900~au, the density must be $>10^3\mathrm{ cm}^{-3}$. The emission spectra calculated by \citet{2002ApJ...576L.149R} include both continuum and line emission in the X-ray band. According to their Equations 8 and 9, a non-radiative shock with $n_0=10^3\mathrm{ cm}^{-3}$ and a diameter equal to the size of the PSF emits just enough flux to explain the X-ray emission from either X-North or X-East. Such a shock would have a mass flux of $\dot M = \mu m_\mathrm{p} n_0 v_\mathrm{sh} \pi (d_\mathrm{sh}/2)^2 \sim 10^{-8}\;M_\sun\;\mathrm{yr}^{-1}$.
Should the densities be higher, both the cooling length and the volume of the shock would be smaller, but the mass flux stays unchanged.

\citet{2020Natur.587..387H} estimate the total mass of the BRN to be $>0.008\;M_\sun{}$. If the mass of the BRN was ejected over the course of 1000~yr, the mass flux is $>10^{-5}\;M_\sun\;\mathrm{yr}^{-1}$ - only 0.1\% of the total mass flux of the BRN outflow is required to power X-North and X-East. In order to generate X-rays here but not in other parts of the BRN, some special condition must trigger a stronger shock here, e.g.\
the initial outflow could be clumpy and inhomogeneous in density and velocity and X-North and X-East represent two locations where such dense clumps are colliding. Given how complex the dynamics of a stellar merger are, it is not unreasonable to expect inhomogeneities in density and speed of factors of a few in the resulting outflow. In this scenario, X-ray sources may flicker on and off over a time scale of $<20$~yr (gas at 500~km~s$^{-1}$ needs 20 years to pass through a cooling zone of $d_{\mathrm{cool}}=2000$~au). This hypothesis is testable with new observations!

The H$\alpha$ emission ring also shows some clumpiness on the outside of the BRN. On the other hand, the UV emission from the BRN itself is remarkably smooth, even when taking into account the larger PSF of \emph{GALEX} ($FWHM \sim 5$~arcsec) compared to the H$\alpha$ observations.

\subsubsection{Shock Interaction with the Circumstellar Medium}
\label{sec:ISM}
If the reverse shock is fast enough, why do we not detect X-rays in the entire region between the forward shock (H$\alpha$) and the FUV emission? In other words, the fact that we see emission only in two specific regions allows us put an upper limit on the density; gas above this density would cause observable X-ray emission as in X-North and X-East.
Assuming the BRN shocks cover a fraction $f_{\Omega} \simeq 0.3$ of the total 4$\pi$ solid angle,  the volume of the hot gas behind the shock is $V_{\rm X} \simeq \frac{R_{\rm BRN}}{4}(4\pi f_{\Omega} R_{\rm BRN}^{3}) \simeq 2\times 10^{56}$ cm$^{-3}$, where we have taken $R_{\rm BRN} \approx 6\times 10^{18}$ cm as the BRN radius, and $R_{\rm BRN}/4$ is the radial thickness of the post-shock gas given its compression by a factor of 4 for a $\gamma = 5/3$ shock.

The X-ray spectrum will be a combination of emission from collisionally excited lines and free-free emission, and for low densities, the plasma may not be in thermal ionization equilibrium after the shock. As an estimate, we continue looking at the free-free emission only. More details on the derivation of the following equations can be found in \citet{1986rpa..book.....R}.

The total free-free X-ray luminosity is given by
\be
L_{\rm X} \simeq j_{\nu}(T_{\rm sh},n_{\rm sh} = 4 n)V_{\rm X},
\ee
where the free-free emissivity of solar-metallicity gas,
\be
j_{\nu} \approx 2\times 10^{-27}T^{1/2}n_{\rm sh}^{2}\bar{g}\,{\rm erg\,s^{-1}\,cm^{-3}},
\ee
$n_{\rm sh}$ is the density of the shocked gas, $n$ is the density of the upstream gas, and $\bar{g} \approx 1.2$ is the Gaunt factor.  Combining, we find
\be
L_{\rm X} \approx 1.2\times 10^{34}\,{\rm erg/s}\left(\frac{n}{1\,{\rm cm^{-3}}}\right)^{2}\left(\frac{v_{\rm sh}}{500\,{\rm km/s}}\right)
\ee
Now, due to the nature of the free-free emission spectrum, the $\nu L_{\nu}$ luminosity in a band $E_X \gtrsim kT$ will be smaller than $L_{\rm X}$ by a factor of $(E_X/kT_{\rm sh})\exp(-E_X/kT_{\rm sh}) \sim 10^{-2}$, where the final numerical estimate assumes $E_X = 2$ keV and $kT_{\rm sh} \sim 0.3$ keV ($v_{\rm sh} \approx 500$ km/s).  Thus, we have
\be
L_{\rm E_X > 1 keV} \sim 10^{-2}L_{\rm X} \sim 10^{32}\,{\rm erg\,s^{-1}}\left(\frac{n}{1\,{\rm cm^{-3}}}\right)^{2},
\ee
which depends sensitively on $kT_{\rm sh}$ (and hence $v_{\rm sh}$).  Thus, given a limit
of a few times $10^{31}$~erg~s$^{-1}$~pc$^{-2}$ (Section~\ref{sec:xraydetection})
we place an upper limit $n = n_{\rm max} \lesssim 1$ cm$^{-3}$ on the upstream density. We note that slower $v_{\rm sh}$ would allow higher densities.

This, in turn, places a limit on the amount of mass in the BRN (or mass of swept-up circumstellar gas in cases when the forward shock dominates),
\be
M_{\rm swept} < n_{\rm max} m_p V_{\rm X} \sim 0.1M_{\odot},
\ee
which is consistent with the lower limit on the BRN mass found by \citet{2020Natur.587..387H} based on the rate of $H_2$ dissociation ($M_{\rm BRN} \gtrsim$ 0.008$M_{\odot}$).

\section{Summary and outlook}
\label{sec:summary}
We detect X-ray emission overlapping the location of TYC 2597-735-1 itself and from two small regions in the outflow. The emission in all three regions is concentrated between about 1 and 2~keV, consistent with a soft, lightly absorbed plasma.  For TYC 2597-735-1 itself, we also present a TESS lightcurve. Despite stray light, the star itself appears to be photometrically variable and this variability is consistent with a period of about 14~days, in-line with that observed previously in optical and infrared RV.

For TYC 2597-735-1, the observed X-ray emission and optical variability can both be explained by coronal activity, which suggests the presence of an efficient convective dynamo in the star. The alternative scenario of X-ray generating shocks in the outflow can also explain the data, but requires either the entire outflow to shock or at least some part of the outflow to be accelerated to twice the escape speed, in contrast to what is observed in other stars that drive comparable outflows. A dynamo is predicted in evolutionary models of post-merger remnant stars and the modeled luminosity is consistent with the observed $\log(L_\mathrm{X}/L_\mathrm{bol})=-5.5$.  We compare the activity in TYC 2597-735-1 with pre-main sequence and main-sequence dwarfs, with single and binary giants, and with other stellar merger candidates. In general, TYC 2597-735-1 exhibits a similar behaviour as other moderately active stars, with an indication that its rotation period is slightly shorter and its Rossby number slightly larger than for other stars of comparable $L_\mathrm{X}/L_\mathrm{bol}$.

As the stellar merger remnant TYC 2597-735-1 continues to cool, contract, and spin-up, we speculate that it may eventually evolve into something akin to an FK Com type star.  This object could therefore provide a missing link connecting this class of rapidly spinning magnetically active stars$-$which, however, no longer exhibit direct evidence for accretion activity or a large-scale bipolar nebula$-$to a past stellar merger event.

We also detect two X-ray sources in the outflow, located between the forward and the reverse shock and hence presumably associated with shock-heated gas. The clumpy nature of the observed X-ray emission might be the result of inhomogeneties in the circumstellar medium or the BRN outflow that lead to significantly higher densities or velocities in these shocked regions, compared to the average density and velocity in the BRN. An upper limit on the total emission from the reverse-shock region constrains the gaseous mass in the BRN to roughly less than a tenth of a solar mass, consistent with previous findings \citep{2020Natur.587..387H} and the expectations of stellar merger simulations given the estimated primary and secondary mass.

The Chandra data presented here were taken serendipitously with an exposure time of less than 10~ks. A deeper observation with Chandra or XMM-Newton can confirm the detection of the X-North and X-East regions and test the hypothesis that regions in the outflow might stochastically become brighter and fade away again on time scales of 20~years; new observations will also reduce the error bars on the spectral properties of TYC 2597-735-1 itself. If the spectrum can be resolved into a low and high-temperature component and the lightcurve is variable, a magnetic corona is required, while the absence of a hotter component and a long-term stable lightcurve indicates an outflow shock. Future instruments with much higher effective area and resolving power are needed to definitively resolve any ambiguity between absorption and soft plasma.

\begin{acknowledgements}
This research has made use of data obtained from the Chandra Data Archive and the Chandra Source Catalog, and software provided by the Chandra X-ray Center (CXC) in the application packages CIAO and Sherpa. This publication makes use of data products from the Two Micron All Sky Survey, which is a joint project of the University of Massachusetts and the Infrared Processing and Analysis Center/California Institute of Technology, funded by the National Aeronautics and Space Administration and the National Science Foundation. 
We acknowledge the use of TESS High Level Science Products (HLSP) produced by the Quick-Look Pipeline (QLP) at the TESS Science Office at MIT, which are publicly available from the Mikulski Archive for Space Telescopes (MAST). Funding for the TESS mission is provided by NASA's Science Mission directorate. This research made use of \texttt{lightkurve}, a Python package for Kepler and TESS data analysis \citep{lightkurve}.
This work has made use of data from the European Space Agency (ESA) mission {\it Gaia} (\url{https://www.cosmos.esa.int/gaia}), processed by the {\it Gaia} Data Processing and Analysis Consortium (DPAC, \url{https://www.cosmos.esa.int/web/gaia/dpac/consortium}). Funding for the DPAC has been provided by national institutions, in particular the institutions participating in the {\it Gaia} Multilateral Agreement.
HMG was supported by the National Aeronautics and Space Administration through Chandra Award Number GO9-20018X issued by the Chandra X-ray Observatory Center, which is operated by the Smithsonian Astrophysical Observatory for and on behalf of the National Aeronautics Space Administration under contract NAS8-03060.
KH was supported by the David {\&} Ellen Lee Prize Postdoctoral Fellowship in Experimental Physics at Caltech and the National Aeronautics and Space Administration Astrophysics Research and Analysis (APRA) Grant Number 80NSSC20K0395.
MNG acknowledges support from MIT's Kavli Institute as a Juan Carlos Torres Fellow and from the European Space Agency (ESA) as an ESA Research Fellow. 
BDM is supported in part by the National Science Foundation (Grant AST-2009255). 
PCS acknowledges support by DLR 50~OR~2102. 
KJS is supported by NASA through the Astrophysics Theory Program (NNX17AG28G and 80NSSC20K0544).
We thank the referee for thoughtful and detailed comments.
\end{acknowledgements}

\facilities{Chandra/ACIS, TESS}

\software{AstroPy \citep{2013A&A...558A..33A,2018AJ....156..123A}, CIAO \citep{2006SPIE.6270E..1VF}, NumPy \citep{van2011numpy,harris2020array}, Matplotlib \citep{Hunter:2007}, Sherpa \citep{2007ASPC..376..543D,doug_burke_2021_4428938}, Lightkurve \citep{lightkurve}, Tesscut \citep{Brasseur2019}, Eleanor \citep{Feinstein2019}}

\bibliography{bib}{}
\bibliographystyle{aasjournal}



\end{document}